\documentclass{article}
\usepackage{spconf,amsmath,graphicx,amssymb,amsfonts}
\usepackage{multirow}
\usepackage{footnote}
\DeclareMathOperator*{\argmax}{argmax}

\title{Reducing Language confusion for Code-switching Speech Recognition with Token-level Language Diarization}
\name{Hexin Liu$^{1,2}$, Haihua Xu$^1$, Leibny Paola Garcia$^3$, Andy~W.~H.~Khong$^2$, Yi He$^1$, Sanjeev Khudanpur$^3$}
\address{$^1$Bytedance AI Lab\\
  $^2$School of Electrical and Electronic Engineering, Nanyang Technological University, Singapore\\
  $^3$CLSP and HLT-COE, Johns Hopkins University, USA}

\begin{document}

\maketitle


\begin{abstract}
 Code-switching~(CS) refers to the phenomenon that languages switch within a speech signal and leads to language confusion for automatic speech recognition~(ASR). This paper aims to address language confusion for improving CS-ASR from two perspectives: incorporating and disentangling language information. We incorporate language information in the CS-ASR model by dynamically biasing the model with token-level language posteriors which are outputs of a sequence-to-sequence auxiliary language diarization module. In contrast, the disentangling process reduces the difference between languages via adversarial training so as to normalize two languages. We conduct the experiments on the SEAME dataset. Compared to the baseline model, both the joint optimization with LD and the language posterior bias achieve performance improvement. The comparison of the proposed methods indicates that incorporating language information is more effective than disentangling for reducing language confusion in CS speech.
\end{abstract}

\section{Introduction}
\label{sec:intro}
Code-switching~(CS) refers to the switching of languages within a spontaneous multilingual recording. Although existing automatic speech recognition~(ASR) methods have shown to achieve good performance on monolingual speech \cite{kaldi,hybrid_ctc_attention_asr}, CS-ASR is still a challenge due to language confusion arising from code switches and the lack of annotated data. 

Language information is often incorporated into CS-ASR models to tackle challenges associated with language confusion. In \cite{zeng19_interspeech}, language identification~(LID) serves as an auxiliary task which enriches the shared encoder with language information. A bi-encoder transformer network was proposed in \cite{bi_encoder}, where two encoders are pre-trained on monolingual data independently to decouple the modeling of Mandarin and English for the capture of language-specific information. This dual-encoder CS-ASR approach has shown to be effective and several methods were subsequently proposed based on this framework \cite{mary20_icassp, song22e_interspeech}. A language-specific attention mechanism has also been proposed to reduce multilingual contextual information for a transformer encoder-decoder CS-ASR model \cite{speech_transformer, zhang22x_interspeech}. In this approach, monolingual token embeddings are separated from code-switching token sequences before being fed into their respective self-attention modules within the decoder layers. 

It is useful to note that the dual-encoder approach, in general, performs LID at frame-level units\textemdash frame-level LID outputs are assigned to the outputs of language-specific encoders before the weighted sum in the mixture-of-experts interpolation process. The frame-level LID, however, is not desirable since the LID performance generally degrades with shorter speech signals \cite{lre17_perform, liu22e_interspeech}. In addition, CS can be
regarded as a speaker-dependent phenomenon \cite{vu2013investigation}, where languages within a CS speech signal share information such as the accent and discourse markers. Therefore, the language-specific attention mechanism would lead to cross-lingual information loss while learning monolingual information. Due to the nature of languages and their transitions in multilingual recordings, exploiting CS approaches at a lower-granularity token level would be more appropriate for CS-ASR.

Language diarization~(LD), as a special case of LID, involves partitioning a code-switching speech signal into homogeneous segments before determining their language identities \cite{Lyu2013LanguageDF, liuxsa}. In our work, LD is reformulated into a sequence-to-sequence task similar to that of ASR to capture token-level language information. Inspired by the success of utterance-level one-hot language vector for multilingual ASR \cite{msr_single, streaming_msr}, we propose to reduce language confusion within CS speech by supplementing the token embeddings with their respective soft language labels\textemdash token-level language posteriors predicted by the LD module\textemdash before feeding these embeddings into the ASR decoder. Since two languages in a CS scenario can be auditorially similar to each other due to the accent and tone of the bilinguist, language posteriors are expected to convey more language information than one-hot language label vectors. Moreover, to explore the effect of language information for CS-ASR, we also propose a second technique to disentangle the language information from the encoder module via adversarial training. The disentangling process aims to normalize two languages so that language confusion resulting from CS can be decreased. Performance evaluations of the proposed methods allow one to gain insights into how language information reduces language confusion for CS-ASR.

\vspace{-0.2cm}
\section{The Hybrid CTC/attention ASR model}
\label{sec:relate}
The hybrid CTC/attention ASR model comprises an encoder module, a decoder module, and a connectionist temporal classification~(CTC) module \cite{hybrid_ctc_attention_asr,ctc}. In our work, the encoder and decoder modules comprise conformer encoder layers and transformer decoder layers \cite{conformer, transformer}, respectively. 

Given a speech signal, we define its acoustic features $\mathbf{X}=(\mathbf{x}_{t} \in \mathbb{R}^{F}| t=1, \ldots , T)$ and token sequence $W=(w_n \in \mathcal{V} | n=1, \ldots , N)$, where $\mathcal{V}$ is a vocabulary of size $V$, $T$ and $N$ are the lengths of acoustic features and token sequence, respectively. These tokens are transformed into $D$-dimensional token embeddings $\mathbf{W}=(\mathbf{w}_{n} \in \mathbb{R}^{D} | n=1, \ldots , N)$ before being fed into the decoder module. The encoder generates $\mathbf{H}=(\mathbf{h}_{t} \in \mathbb{R}^{D}| t=1, \ldots , T_{1})$ from inputs $\mathbf{X}$, which are then used as inputs for the decoder and CTC modules. During training, masked self-attention of the token embeddings and the cross-attention between hidden output and token embeddings are performed within the decoder layer. Here, the masked self-attention process prevents the current tokens from attending to future tokens. The decoder predicts the next token $w_{n}$ based on historical tokens $w_{1:n-1}$ and $\mathbf{H}$ via
\begin{equation}
\setlength{\abovedisplayskip}{4pt}
\setlength{\belowdisplayskip}{4pt}
  p\left(w_{n}|w_{1:n-1},\mathbf{X}\right)=\mathrm{Decoder}\left ( \mathbf{w}_{1:n-1}, \mathbf{H} \right ),
  \label{eq:asr_decoder}
\end{equation}
where $p(w_{n}|w_{1:n-1},\mathbf{X})$ is the posterior of decoding $w_{n}$ given acoustic features and historical tokens, and $\mathrm{Decoder}(\cdot)$ denotes the transformer-based ASR decoder. The model is optimized via a multi-task objective function
\begin{equation}
\setlength{\abovedisplayskip}{4pt}
\setlength{\belowdisplayskip}{4pt}
  \mathcal{L}_{\mathrm{asr}}=\alpha \mathcal{L}_{\mathrm{ctc}} + \left ( 1-\alpha  \right ) \mathcal{L}_{\mathrm{att}},
  \label{eq:loss_asr}
\end{equation}
where $\mathcal{L}_{\mathrm{ctc}} $ and $\mathcal{L}_{\mathrm{att}}$ denote the CTC loss and encoder-decoder cross-entropy loss with label smoothing, respectively, and $\alpha$ is a parameter associated with multi-task learning. The decoding process is defined to maximize the linear combination of the logarithmic CTC and attention objectives
\begin{equation}
\small
\setlength{\abovedisplayskip}{4pt}
\setlength{\belowdisplayskip}{4pt}
  \widehat{W} = \argmax_W \big \{ \alpha \mathrm{log} p_{\mathrm{ctc}}\left(W|\mathbf{X}\right) + \left ( 1-\alpha  \right ) \mathrm{log} p_{\mathrm{att}}\left(W|\mathbf{X}\right)  \big \}.
  \label{eq:decoding}
\end{equation}

\vspace{-0.2cm}
\section{Reducing language confusion}
\label{sec:method}
\begin{figure}[t]
\setlength{\belowcaptionskip}{-1cm}
  \centering
  \includegraphics[width=\linewidth]{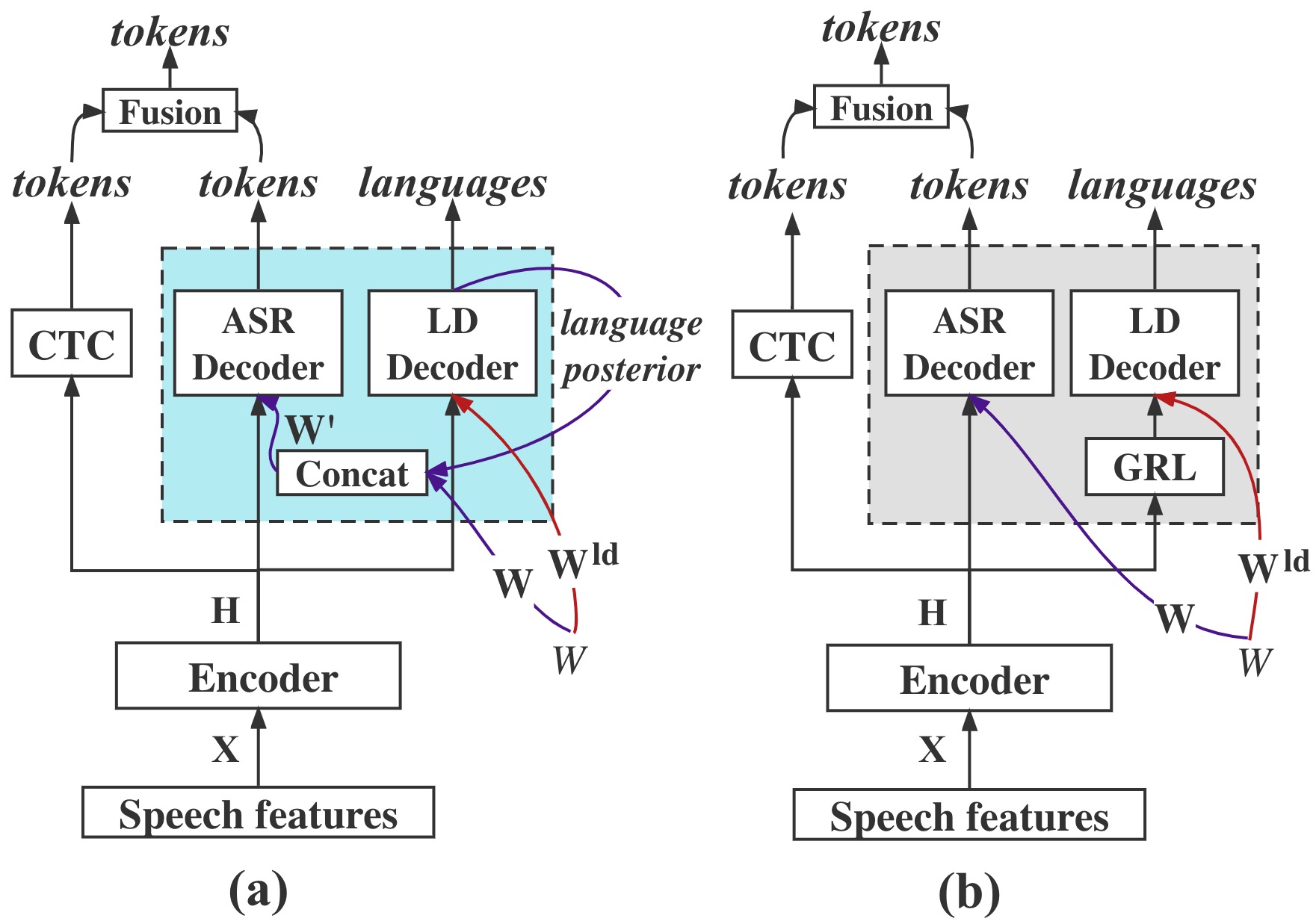}
  \caption{The hybrid CTC/attention model with (a) incorporating language information using language posterior bias, and (b) disentangling language via adversarial training.}
  \label{fig:joint_asr_ld}
\end{figure}

\vspace{-0.2cm}
\subsection{Sequence-to-sequence language diarization}
\label{sec:joint}
As shown in Fig.~\ref{fig:joint_asr_ld}, given the encoder outputs $\mathbf{H}$ and the token sequence $W$ with the corresponding token embeddings $\mathbf{W}^{\mathrm{ld}}=(\mathbf{w}^{\mathrm{ld}}_{n} \in \mathbb{R}^{D} | n=1, \ldots , N)$, we propose to decode the token-level language labels $L=(l_n \in \mathcal{V}^{\mathrm{ld}}| n=1, \ldots , N)$ of $W$ via a transformer decoder being denoted as LD decoder in the proposed model. Here, $\mathcal{V}^{\mathrm{ld}}$ denotes the language vocabulary. 

While tokens within a sequence possess contextual information for the ASR task, they share the language identities in the LD task, i.e., they should either be of the same or different languages. Such information is expected to be available in both self-attention and cross-attention processes. In addition, masking out future tokens is required when training the ASR decoder to minimize the mismatch between training and decoding. Masking future tokens during training, however, results in a lower triangular self-attention weight matrix leading to language-related information loss. We, therefore, propose to utilize the complete token sequence to train the LD decoder during the self-attention process if the LD decoder participates in only multi-task optimization but not the decoding process. To this end, we have
\begin{equation}
\setlength{\abovedisplayskip}{4pt}
\setlength{\belowdisplayskip}{4pt}
  p\left(l_{n}|w_{1:N},\mathbf{X}\right)=\mathrm{Decoder}^{\mathrm{ld}}\left ( \mathbf{w}^{\mathrm{ld}}_{1:N}, \mathbf{H} \right ),
  \label{eq:ld_decoder}
\end{equation}
where $\mathrm{Decoder}^{\mathrm{ld}}(\cdot)$ denotes the LD decoder. On the other hand, if the LD decoder participates in the decoding phase, future context should be masked to reduce the mismatch between training and decoding. The LD decoder is then jointly optimized with the hybrid CTC/attention CS-ASR model via
\begin{equation}
\setlength{\abovedisplayskip}{4pt}
\setlength{\belowdisplayskip}{4pt}
  \mathcal{L}_{\mathrm{joint}}=\alpha \mathcal{L}_{\mathrm{ctc}} + \left ( 1-\alpha  \right ) \mathcal{L}_{\mathrm{att}} + \beta \mathcal{L}_{\mathrm{ld}},
  \label{eq:loss_joint}
\end{equation}
where $\beta$ is a multi-task learning parameter, and $\mathcal{L}_{{\mathrm{ld}}}$ is a label-smoothed cross-entropy loss between the predicted and ground-truth language labels for the LD decoder.

\vspace{-0.2cm}
\subsection{Language posterior bias}
\label{sec:language_bias}
The utterance-level language vector has shown to be effective for multilingual ASR \cite{msr_single, streaming_msr}. Since intra-sentence code-switching occurs at word level, we propose to bias the ASR output using token-level language posteriors predicted by the LD decoder as shown in Fig.~\ref{fig:joint_asr_ld}~(a). 

Given historical tokens $w_{1:n-1}$ and hidden output $\mathbf{H}$, the LD decoder generates a $V^{ld}$-dimensional language posterior vector $\mathbf{p}(l_{n-1}|w_{1:n-1},\mathbf{X})$ after a softmax function. Token embedding $\mathbf{w}_{n-1}$ is subsequently concatenated with its language posterior vector $\mathbf{p}$ before being fed into the ASR decoder. The ASR decoder output is next computed via
\begin{eqnarray}
\setlength{\abovedisplayskip}{4pt}
\setlength{\belowdisplayskip}{4pt}
  \mathbf{w}^{\prime}_{n-1} = \mathrm{Concat}\big (\mathbf{w}_{n-1}, \mathbf{p}\left(l_{n-1}|w_{1:n-1},\mathbf{X}\right) \big ),
  \label{eq:w_prime} 
  \\
  p\left(w_{n}|w_{1:n-1},\mathbf{X}\right) = \mathrm{Decoder}\left ( \mathbf{w}^{\prime}_{1:n-1}, \mathbf{H} \right ),
  \label{eq:ctc}
\end{eqnarray}
where $\mathrm{Concat}(\cdot)$ denotes the concatenation operation, and $\mathbf{W}^{\prime}=(\mathbf{w}^{\prime}_{n} \in \mathbb{R}^{D+V^{\mathrm{ld}}} | n=1, \ldots , N)$ are input token embeddings of the ASR decoder which are subsequently projected back to $D$ dimensions by a linear layer. The model is optimized via (\ref{eq:loss_joint}) and the decoding process is similar to (\ref{eq:decoding}) but the input token embeddings $\mathbf{W}$ of the ASR decoder are replaced with $\mathbf{W}^{\prime}$.
\vspace{-0.2cm}
\subsection{Disentangling with adversarial training}
\label{sec:disentangle}
In contrast to incorporating language information, disentangling language information aims to normalize the two languages by reducing the difference between them. To disentangle language information from the encoder, we employ a gradient reversal layer~(GRL) between the encoder and LD decoder as shown in Fig.~\ref{fig:joint_asr_ld}(b). The GRL was originally proposed for unsupervised domain adaptation in \cite{ganin2015unsupervised}\textemdash it remains in the forward process while reversing the gradient during backpropagation. In our CS-ASR context, the model achieves adversarial training that reduces the language information within the encoder while optimizing the LD decoder to be discriminative.
\vspace{-0.2cm}
\section{Dataset, Experiment, and Results}
\label{sec:experiment}
\vspace{-0.2cm}
\subsection{Dataset and experiment setup}
We conducted all experiments on the SEAME dataset\textemdash a Mandarin-English code-switching corpus containing spontaneous conversational speech \cite{seame}. Both intra- and inter-sentence code-switching speech exist in this dataset. We divide this dataset into a 98.02-hour training set, a 5.09-hour validation set, and two test sets $\textit{dev}_{\texttt{man}}$ and $\textit{dev}_{\texttt{sge}}$ in the same manner as \cite{zeng19_interspeech}. Details of the test sets are presented in Table~{\ref{tab:data}. 

We implemented the models and evaluated them on ESPnet~\footnote{Source code: https://github.com/Lhx94As/reducing\_language\_confusion} \cite{espnet}. During training and inference, we followed the experimental setup of the espnet2 SEAME recipe in regard to data preprocessing and model configuration. Speech perturbation (with factors 0.9, 1.0, and 1.1) and SpecAugment are applied as data augmentation \cite{pertub, specaug}. Words are transformed into a total of $V=5,628$ tokens including 3,000 English byte-pair encoding~(BPE) tokens, 2,624 Mandarin characters, and 4 special tokens for {\em unk}, {\em noise}, {\em blank}, and {\em sos/eos}. All tokens are transformed to build $\mathcal{V}^{\mathrm{ld}}$ comprising {\em e} for English BPEs, {\em m} for Mandarin characters, {\em sos/eos}, and {\em other} for other tokens. Tokens in $\mathcal{V}^{\mathrm{ld}}$ are used as LD outputs. We extracted $F=83$ dimensional features comprising 80-dimensional log-fbanks and 3-dimensional pitch for each speech segment before applying global mean and variance normalization.

The baseline model is a hybrid CTC/Attention ASR model comprising twelve conformer encoder layers and six transformer decoder layers \cite{conformer, transformer, karita19_interspeech}. The proposed LD decoder employs the same configuration as the ASR decoder. All self-attention encoder and decoder layers have four attention heads with input and output dimensions being $D=256$, and the inner layer of the position-wise feed-forward network is of dimensionality 2048. The language model~(LM) is a sixteen-layer transformer model, where each attention layer has eight heads. Parameters $\alpha=0.3$ and a label smoothing factor of 0.1 is used in (\ref{eq:loss_asr}). All models are trained for 50 epochs on two V100 GPUs. During inference, the 10-best models during validation are averaged. We adopted the 10-best beam search method with $\alpha=0.4$ in (\ref{eq:decoding}). The proposed systems are evaluated by employing mix error rate~(MER) comprising word error rate~(WER) for English and character error rate~(CER) for Mandarin.

\begin{table}[t]
\centering
\caption{Details of two test sets divided from SEAME}
\label{tab:data}
\setlength{\tabcolsep}{2mm}{
\begin{tabular}{c|c|c|ccc}
\hline
\multirow{2}{*}{} & \multirow{2}{*}{\textbf{Speakers}} & \multirow{2}{*}{\textbf{Hours}} & \multicolumn{3}{c}{\textbf{Duration ratio (\%)}}                 \\ \cline{4-6} 
         &       &           & \multicolumn{1}{c|}{Man} & \multicolumn{1}{c|}{Eng} & CS \\ \hline
$\textit{dev}_{\texttt{man}}$    & 10    & 7.49      & \multicolumn{1}{c|}{14}  & \multicolumn{1}{c|}{7}   & 79 \\ \hline
$\textit{dev}_{\texttt{sge}}$    & 10    & 3.93      & \multicolumn{1}{c|}{6}   & \multicolumn{1}{c|}{41}  & 53 \\ \hline
\end{tabular}}
\vspace{-0.2cm}
\end{table}

\vspace{-0.2cm}
\subsection{Results of jointly optimizing CS-ASR and LD}
\label{sec:result_joint}

\begin{figure}[t]
\setlength{\abovecaptionskip}{-0.4cm}
  \centering
  \includegraphics[width=\linewidth]{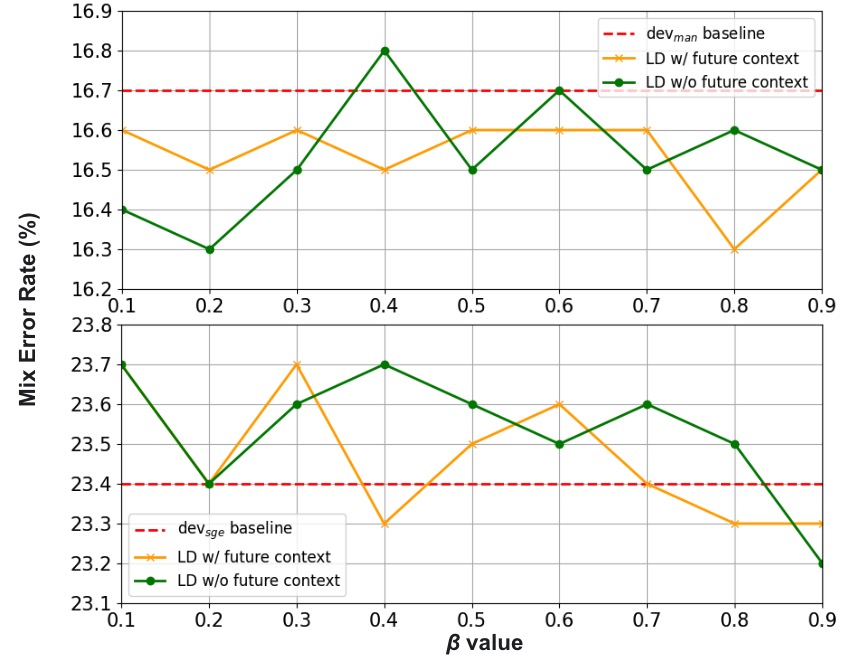}
  \caption{Performance evaluation of the CS-ASR-LD models with different $\beta$ values by employing MER (\%), where orange and green lines denote training the LD decoders with and without future context, respectively.}
  \label{fig:results_cs_asr_ld}
\vspace{-0.2cm}
\end{figure}
We denote the hybrid CTC/attention CS-ASR model which is jointly trained with LD as CS-ASR-LD. The CS-ASR-LD model is evaluated on $\textit{dev}_{\texttt{man}}$ and $\textit{dev}_{\texttt{sge}}$ to explore the effect of the auxiliary LD task with different $\beta$ values. In addition, we compare the CS-ASR performance with different LD decoders trained on tokens with and without future context, respectively. The above results are present in Fig.~\ref{fig:results_cs_asr_ld}. 

As opposed to the CS-ASR-LD model with an LD decoder that masks future context during training, those with the LD decoder trained on full context consistently exhibit lower MER on $\textit{dev}_{\texttt{man}}$ than the baseline. In addition, the CS-ASR-LD model whose LD decoder does not mask future context during training achieves moderate performance improvement on $\textit{dev}_{\texttt{sge}}$ with high $\beta$ values compared to the baseline. The best overall performance of 16.3\% and 23.3\% on $\textit{dev}_{\texttt{man}}$ and $\textit{dev}_{\texttt{sge}}$, respectively, is achieved when training the LD decoder on full context with $\beta$=0.8.
Since $\textit{dev}_{\texttt{man}}$ contains mostly CS speech, the above results indicate that incorporating discriminative language information benefits ASR performance on CS speech.

\vspace{-0.2cm}
\subsection{Results of biasing CS-ASR with language posterior}
We evaluate the hybrid CTC/attention model with language posterior bias~(LPB), which is the LD decoder output, on $\textit{dev}_{\texttt{man}}$ and $\textit{dev}_{\texttt{sge}}$. These results and ablation studies are shown in Table~\ref{tab:language_bias} with $\beta=0.8$. Here, employing LPB without joint optimization with LD is achieved by intercepting the backpropagation from the LD decoder to the encoder. 

Compared with the baseline, employing either LD or LPB exhibits better performance on $\textit{dev}_{\texttt{man}}$ which contains mainly CS test speech. Systems S2.4 and S2.5 suffer from moderate performance degradation on $\textit{dev}_{\texttt{sge}}$. Since $\textit{dev}_{\texttt{sge}}$ comprises more monolingual speech where less language confusion exists compared to $\textit{dev}_{\texttt{man}}$, the above results imply that utilizing token-level language information, while beneficial for CS-ASR, may not contribute effectively to monolingual ASR.

System S2.8 with both LD and LPB achieves the best performance of 16.3\% and 23.0\% on $\textit{dev}_{\texttt{man}}$ and $\textit{dev}_{\texttt{sge}}$, respectively, among all systems that do not employ LM. This indicates the efficacy of our proposed method. In addition, while employing LM improves the CS-ASR performance, the proposed language-aware methods do not benefit from the use of LM as much compared to the baseline model. This also implies that a language-aware LM would be more appropriate to CS-ASR compared to the general LM \cite{syntactic_LM}.

\begin{table}[t]
\centering
\caption{Performance comparison between baseline model and those with language posterior bias~(LPB) by employing MER~(\%), ``Future context'' denotes whether the future tokens are used or not for the LD decoder during training}
\label{tab:language_bias}
\setlength{\tabcolsep}{1.45mm}{
\begin{tabular}{c|c|c|c|c|cc}
\hline
\multirow{2}{*}{\textbf{Index}} & \multirow{2}{*}{\textbf{Model}} & \multirow{2}{*}{$\beta$} & \multirow{2}{*}{\begin{tabular}[c]{@{}c@{}}\textbf{Future}\\ \textbf{context}\end{tabular}} & \multirow{2}{*}{\textbf{LM}} & \multicolumn{2}{c}{\textbf{MER}}              \\ \cline{6-7} 
&     &   &  & &\multicolumn{1}{c|}{$\textit{dev}_{\texttt{man}}$} & $\textit{dev}_{\texttt{sge}}$ \\ \hline
S0    & \multirow{2}{*}{baseline}   & -   & - & No  & \multicolumn{1}{c|}{16.7}   & 23.4   \\ \cline{1-1} \cline{3-7} 
S1    &    & -   & - & Yes & \multicolumn{1}{c|}{16.5}   & 23.0   \\ \hline
S2.1  & \multirow{3}{*}{+LD}   & 0.8   & Yes & No & \multicolumn{1}{c|}{16.3}   & 23.3   \\ \cline{1-1} \cline{3-7} 
S2.2  &    & 0.8   & No & No & \multicolumn{1}{c|}{16.6}   & 23.5   \\ \cline{1-1} \cline{3-7} 
S2.3  &    & 0.8   & Yes & Yes & \multicolumn{1}{c|}{16.2}   & 23.1  \\ \hline
S2.4  & \multirow{3}{*}{+LPB}  & 0.8   & Yes & No & \multicolumn{1}{c|}{16.4}   & 23.5   \\ \cline{1-1} \cline{3-7}
S2.5  &    & 0.8   & No & No & \multicolumn{1}{c|}{16.6}   & 23.5   \\ \cline{1-1} \cline{3-7} 
S2.6  &    & 0.8   & No & Yes & \multicolumn{1}{c|}{16.4}   & 23.2   \\ \hline
S2.7  & \multirow{4}{*}{+LD +LPB}   & 0.8  & Yes & No & \multicolumn{1}{c|}{16.6}   & 23.4  \\ \cline{1-1} \cline{3-7}
S2.8  &    & 0.8   & No & No & \multicolumn{1}{c|}{16.3}   & 23.0   \\ \cline{1-1} \cline{3-7} 
S2.9  &    & 0.8   & Yes & Yes & \multicolumn{1}{c|}{16.5}   & 22.9  \\ \cline{1-1} \cline{3-7} 
S2.10 &    & 0.8   & No & Yes & \multicolumn{1}{c|}{16.1}   & 22.8  \\ \hline
\end{tabular}}
\vspace{-0.2cm}
\end{table}

\vspace{-0.2cm}
\subsection{Results of disentangling language information}
Language information is disentangled from the CS-ASR model via adversarial training with a GRL. We evaluate this approach with $\beta=0.2, 0.5,$ and 0.8 and results are shown in Table~\ref{tab:grl}.

Disentangling language information from the CS-ASR model (while maintaining the MER on $\textit{dev}_{\texttt{man}}$) results in performance degradation on $\textit{dev}_{\texttt{sge}}$ compared to the baseline. Since $\textit{dev}_{\texttt{sge}}$ contains more monolingual speech than $\textit{dev}_{\texttt{man}}$, these results suggest that language information contributes to the CS-ASR performance even for monolingual speech. This is because the process of ASR intrinsically encodes language-related information such as phonotactics and syntax into the model \cite{efficient_lid}. Reducing the language information, therefore, may degrade the ASR performance on monolingual speech.

\begin{table}[t]
\centering
\caption{Performance comparison between baseline model and those after disentangling language information by employing MER~(\%), ``Future context'' denotes whether the future tokens are used or not for the LD decoder during training}
\label{tab:grl}
\setlength{\tabcolsep}{1.7mm}{
\begin{tabular}{c|c|c|c|cc}
\hline
\multirow{2}{*}{\textbf{Index}} & \multirow{2}{*}{\textbf{Model}} & \multirow{2}{*}{$\beta$} & \multirow{2}{*}{\begin{tabular}[c]{@{}c@{}}\textbf{Future}\\ \textbf{context}\end{tabular}} & \multicolumn{2}{c}{\textbf{MER}}              \\ \cline{5-6} 
&   &   &   &\multicolumn{1}{c|}{$\textit{dev}_{\texttt{man}}$} & $\textit{dev}_{\texttt{sge}}$ \\ \hline
S0  & baseline   & -   & -   & \multicolumn{1}{c|}{16.7}   & 23.4   \\ \hline
S3.1  & \multirow{3}{*}{+GRL}   & 0.2    & Yes  & \multicolumn{1}{c|}{16.8}   & 23.6   \\ \cline{1-1} \cline{3-6}
S3.2  &    & 0.5   & Yes  & \multicolumn{1}{c|}{16.7}   & 23.7  \\ \cline{1-1} \cline{3-6}
S3.2  &    & 0.8   & Yes  & \multicolumn{1}{c|}{16.7}   & 23.7   \\ \hline
\end{tabular}}
\end{table}

\section{Conclusion}
\label{sec:conclud}
We proposed two methods to reduce language confusion for CS-ASR. The first method is to incorporate language information using the language posterior bias, while the second method is to disentangle language information via adversarial training. Compared to the baseline model, our proposed LPB method in conjunction with multi-task learning exhibits higher performance on the SEAME dataset. The comparison between the proposed LPB method and the adversarial training for CS-ASR-LD highlights the efficacy of incorporating language information to improve CS-ASR. In addition, the proposed language-aware CS-ASR models achieve lower performance improvement after employing LM during decoding compared to the baseline model. This indicates that general LM may not be suitable for language-aware CS-ASR models due to the lack of language information.

\vfill\pagebreak
\small
\bibliographystyle{IEEEtran}
\bibliography{strings, refs}

\end{document}